\documentclass[aps,prb,twocolumn,groupedaddress,showpacs]{revtex4}

\usepackage[bookmarks]{hyperref}
\usepackage[dvips]{graphicx}
\usepackage{amsmath}
\usepackage{amstext}
\usepackage{graphics}
\usepackage{exscale}
\usepackage{epsfig}
\usepackage[T1]{fontenc}
\usepackage{bm}
\usepackage{bbm}
\usepackage{color}

\usepackage{version}

\usepackage{latexsym}

\usepackage[tight]{subfigure}

\renewcommand{\epsilon}{\varepsilon}

\newcommand{\bra}[1]{ \langle \! #1\ \! |} 
\newcommand{\ket}[1]{\! | #1\ \!\! \rangle}

\newcommand{\elcre}[2]{ c^{\dagger}_{#1,#2}}
\newcommand{\elann}[2]{ c_{#1,#2}}

\newcommand{\e}{\mathrm e}

\newcommand{\vct}[1]{\bm #1}
\newcommand{\vk}{{\bm k}}

\newcommand{\hc}{\mathrm{h.c.}}

\includeversion{commentst1} 
\begin{document}

\title{Microscopic resolution of the interplay of Kondo screening
  and superconducting pairing}  
\author{Johannes Bauer${}^{1,2}$, Jose I. Pascual${}^{3,4}$, and Katharina J. Franke${}^{3,5}$}
\affiliation{${}^1$Max-Planck Institute for Solid State Research,
Heisenbergstr. 1, 70569 Stuttgart, Germany} 
\affiliation{${}^2$Department of Physics, Harvard University, Cambridge,
  Massachusetts 02138, USA}
\affiliation{${}^3$  Freie Universit\"at Berlin, Institut f\"ur
  Experimentalphysik, Arnimallee 14, 14195 Berlin, Germany}
\affiliation{${}^4$ CIC nanoGUNE, 20018 Donostia-San Sebastian, and
  Ikerbasque, Basque Foundation for Science, 48011 Bilbao, Spain}
\affiliation{${}^5$  Technische Universit\"at Berlin, Institut f\"ur Festk\"orperphysik, Hardenbergstra\ss e 36, 10623 Berlin, Germany}
\date{\today} 

\begin{abstract}
Magnetic molecules adsorbed on a superconductor give rise to a local
competition of Cooper pair and Kondo singlet formation inducing subgap bound
states.
For Manganese-phthalocyanine molecules on a Pb(111) substrate,
scanning tunneling spectroscopy resolves pairs of subgap bound states and
two Kondo screening channels. We show in a combined approach of scaling
and numerical renormalization group calculations that the intriguing
relation between Kondo screening and superconducting pairing is solely
determined by the hybridization strength with the substrate. We demonstrate
that an effective one-channel Anderson impurity model with a sizable 
particle-hole asymmetry captures universal and non-universal observations in
the system quantitatively. The model parameters and disentanglement of the two
screening channels are elucidated by scaling arguments.   
\end{abstract}
% 72.10.Fk Scattering by point defects, dislocations, surfaces, and other imperfections (including Kondo effect) 
%72.15.Qm Scattering mechanisms and Kondo effect (see also 75.20.Hr Local moments in compounds and alloys; Kondo effect, valence fluctuations, heavy fermions in magnetic properties and materials)
%75.20.Hr Local moment in compounds and alloys; Kondo effect, valence fluctuations, heavy fermions (see also 72.15.Qm Scattering mechanisms and Kondo effect)
%73.20.-r Electron states at surfaces and interfaces
%73.20.Hb Impurity and defect levels; energy states of adsorbed species 
%74.81.-g Inhomogeneous superconductors and superconducting systems 

\maketitle

%\paragraph{Introduction -} 
\section{Introduction}
Metals become superconducting, when their electrons form singlet Cooper pairs
via an attractive interaction. On the other hand, electrons in metals can also
undergo another type of singlet formation, namely to form a Kondo
screening cloud, when magnetic impurities are present. 
The fascinating interplay of Cooper pair and
Kondo singlet formation
\cite{AG61,ZM70,Zit70,MZ71,Shi73,Mat77,Sak70,BVZ06,Kon64,hewson} can be
microscopically observed when magnetic atoms or molecules are adsorbed on superconducting
surfaces.  
The ground state of such a combined system has been predicted to be either a
Kondo screened singlet state ($S=0$), if the Kondo scale $k_{\rm B}\, T_{\rm
  K}$ is much larger than the superconducting pairing energy
$\Delta_{\rm sc}$, or an unscreened multiplet state ($S>0$) for
$k_{\rm B}\, T_{\rm   K}\ll \Delta_{\rm  sc}$. Characteristic
features include subgap states with bound state energies $E_b$ - often called Shiba states. 
For the singlet ground state with screened impurity spin, the bound states are $S>0$
excitations at $E_b < 0$, where the Kondo singlet is broken, and for the multiplet
ground state the bound states with $E_b > 0$ are singlet excitations including Kondo screening.
When the energy scale for Kondo singlet formation becomes smaller and
comparable with the Cooper pairing energy, $k_{\rm  B}\, T_{\rm
  K}\sim \Delta_{\rm  sc}$, the bound state energies at $E_b$ go
to $0$, and the $S>0$ state becomes the ground state. At this point a 
quantum phase transition (QPT) occurs. 
The Cooper pair breaking effect is expected to behave like $
1-(E_b/\Delta_{\rm sc})^2$ and thus is 
 most effective here leading to a strong suppression of the
superconducting $T_c$ for larger impurity concentrations.\cite{MZ71,Shi73,BVZ06}

For a long time the accurate resolution of the subgap bound states and their
dependence on $T_{\rm   K}$ and $\Delta_{\rm  sc}$ have remained elusive. 
Recently, the bound states have been analyzed in tunable
mesoscopic superconductor-quantum dot-normal lead structures.\cite{Dea10a}
Scanning tunneling microscopy (STM) has been used to 
detect the local influence of single magnetic atoms on a
superconducting substrate.\cite{YJLCE97}
To study the interplay, an experiment with variable
magnetic interaction strengths is desirable. Manganese-phthalocyanine
(MnPc) molecules on a Pb(111) substrate form 
a Moire-like superstructure. 
Tunneling spectroscopy on different Mn sites reveals two Kondo screening
channels and pairs of bound states of varying energy $E_b$.
The large number of different adsorption sites leads to a variety of magnetic
interactions and Kondo scales.
The smaller Kondo scale lies in the interesting regime $k_{\rm  B}\, T_{\rm
  K}\sim \Delta_{\rm  sc}$. 
Shiba states crossing the Fermi level and the predicted QPT could be
observed.\cite{FSP11} A number of important question remained however
unresolved. 
What is the role of the second Kondo screening channel and does it give rise
to a shift of the critical point of the QPT? Which microscopic parameter
drives the behavior of the system across the QPT? Does the asymmetry in the
STM intensity of the bound states reveal particular physical properties of the system? 

Our theoretical approach to address those questions is a combination of scaling
arguments and numerical renormalization group (NRG) calculations. 
\cite{BCP08,SSSS92,SSSS93,YO00,BOH07}
The former are used to connect the complex experimental situation to an 
effective Anderson impurity model with one relevant channel and its model parameters.
NRG calculations for this model demonstrate that the experimental behavior 
can be quantitatively understood by {\em only}
varying the hybridization between MnPc and the substrate. 
The results are reliable as the NRG is known to capture the Kondo effect
accurately in contrast to many other methods, used to describe impurities in
superconductors, which contain mean field aspects like classical
spins.\cite{ZM70,Zit70,MZ71,Shi73,Mat77,FB97a,FB97b,SBS97}
The accuracy of the theoretical modeling is tested by the direct comparison of the
experimental and theoretical results for the point of the QPT and the
positions and weights of the bound states.     

\begin{figure}[htbp]
%\vspace*{-0.5cm}
\centering
\includegraphics[width=0.95\columnwidth]{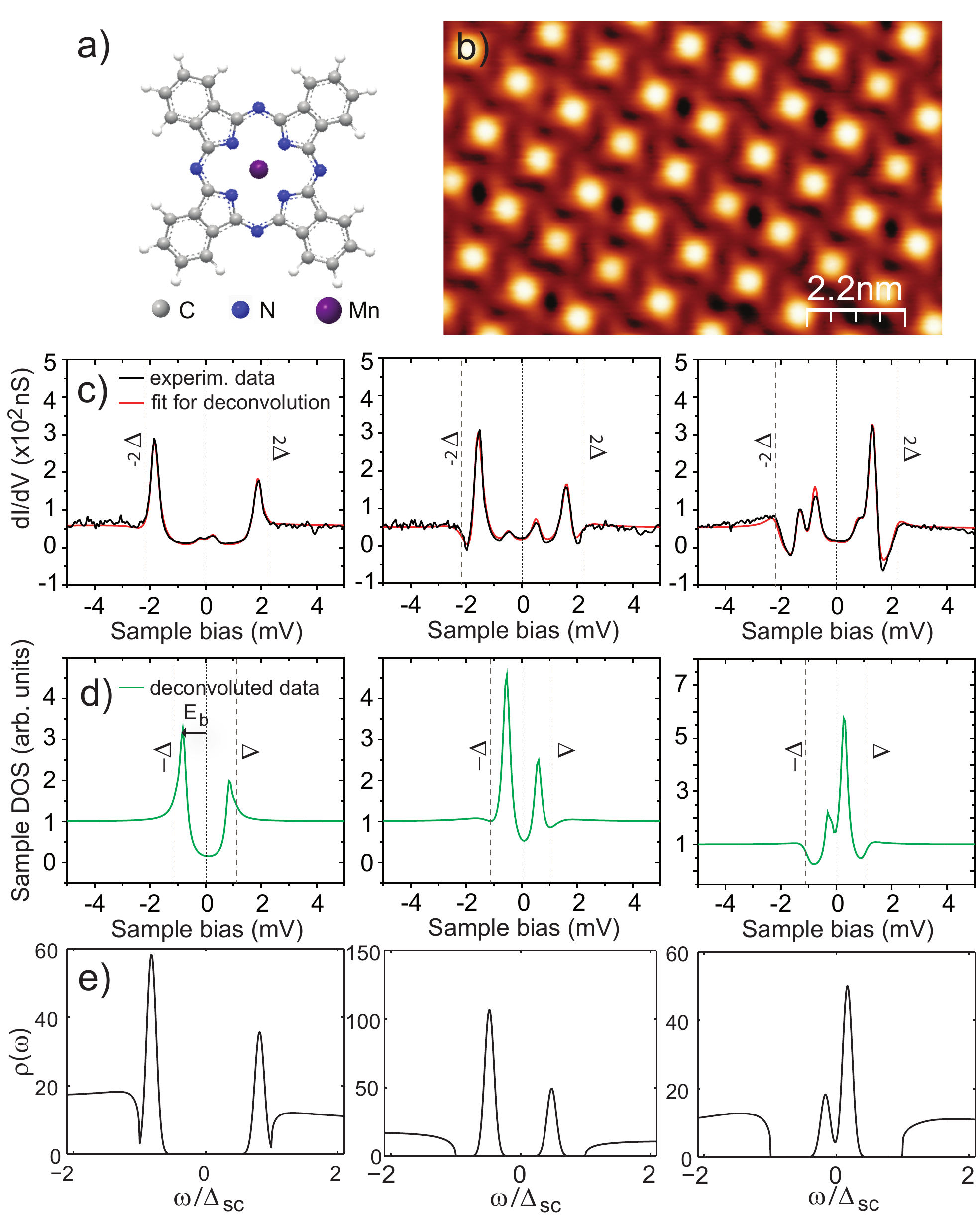}
\vspace*{-0.5cm}
%\vspace*{-0.3cm}
\caption{(Color online) a) Molecular structure of
  Manganese-Phthalocyanine (MnPc), b) Constant-current STM image of
  highly ordered island of MnPc on Pb(111) (I\,=\,23\,pA,
  $V$\,=\,100\,mV), c) Tunneling spectra taken on different molecules
  on top of the Mn center (feedback opened at I\,=\,450\,pA,
  $V$\,=\,8.6\,mV). Also shown are the fits used to
  derive the deconvoluted spectra below in (d). 
  e) Examples of NRG spectra with broadened subgap peaks for similar
  $E_{b,\alpha}$.}           
\label{fig:1}
\vspace*{-0.5cm}
\end{figure}

%\paragraph{Experimental details -} 
\section{Experimental results}
As detailed in Ref.~\onlinecite{FSP11}, the MnPc molecules (see Fig.~\ref{fig:1}(a))
have been deposited on an atomically clean Pb(111) 
substrate at room temperature under ultra-high vacuum
conditions. STM at 4.5\,K resolves highly
ordered islands (Fig.~\ref{fig:1}(b)). Tunneling spectroscopy has been
used to resolve the superconducting gap structure and its subgap
states, as well as Kondo resonances on different molecules.  
The superconducting state of the Pb(111) substrate and Pb tip shows as
pronounced differential conductance peaks at $E=\pm2\Delta_{\rm sc}$
with $\Delta_{\rm sc}=1.1$\,meV.

\begin{figure}[htbp]
%\vspace*{1cm}
\vspace*{-0.5cm}
\centering
\includegraphics[width=0.95\columnwidth]{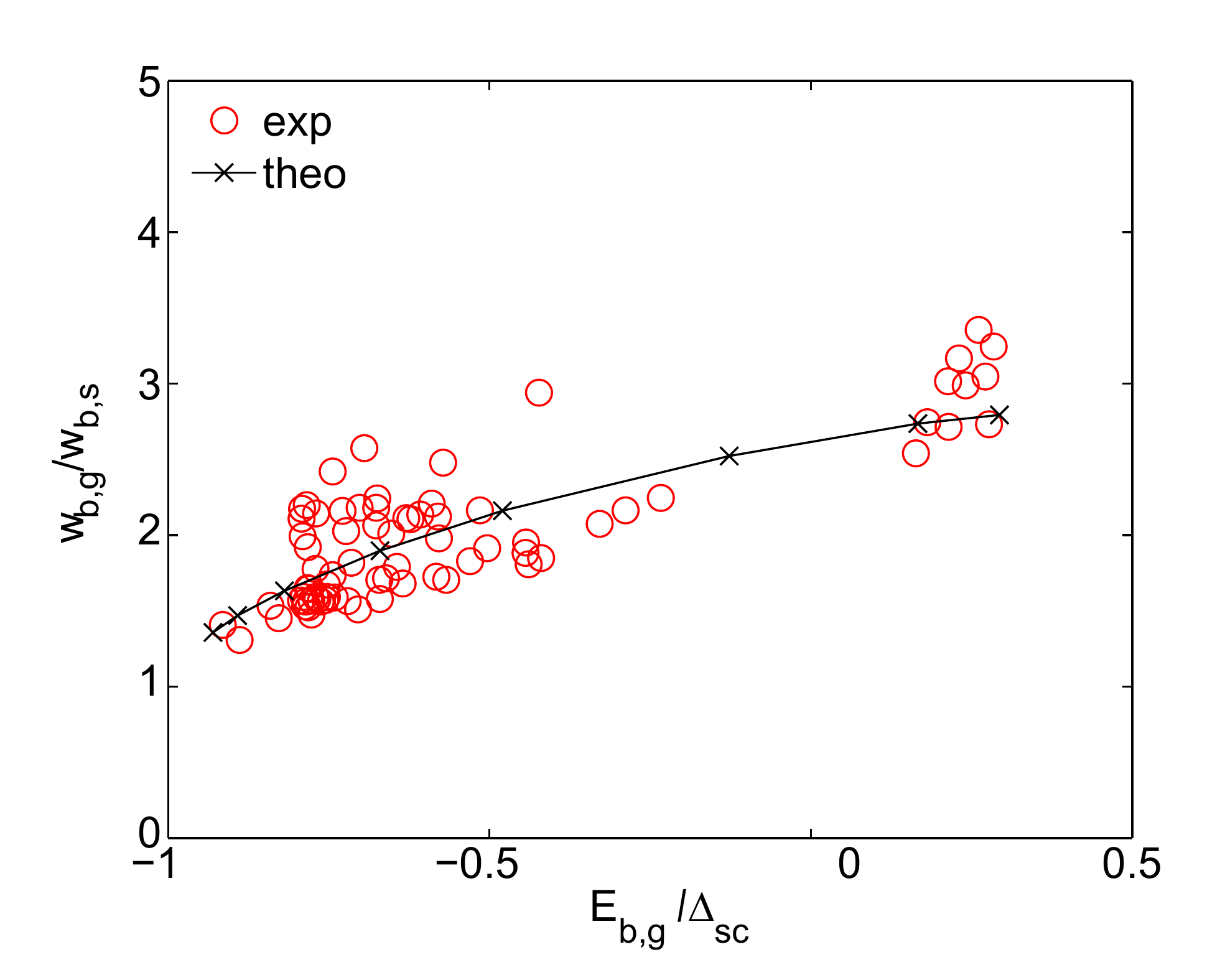}
\vspace*{-0.5cm}
%\includegraphics[width=0.45\textwidth]{wbgowbsvsEb_expdata_comptheo.eps}
%\vspace*{-0.3cm}
\caption{(Color online) Ratio of bound state weights
  $w_{b,g}/w_{b,s}$ vs bound state energy $E_{b,g}/\Delta_{\rm
    sc}$. Comparison of experimental results with theory. In all the
  range of bound state energy we used the model parameters  $\epsilon_{\rm
    d}/\Delta_{\rm sc}\approx -73$, and $U/\Delta_{\rm sc}\approx 91$
  and the overlap $\Gamma$ was varied.}  
\label{fig:wbgowbs}
\end{figure}

%\paragraph{Experimental results -} 
Three differential conductance spectra taken on different MnPc
molecules are shown in Fig.~\ref{fig:1}(c). 
Two larger peaks as well as the two smaller peaks are located at
symmetric bias voltages within the gap of the superconductor-superconductor
tunneling barrier. The larger peaks are an expression of the Shiba states and
indicate the magnetic interaction with the superconducting substrate. The
smaller peaks are a result of thermal excitations at the measurement
temperature of 4.5\,K across the gap.  
In order to remove the effect of the superconducting tip and finite
temperature on the tunneling spectra, we developed a deconvolution method
\cite{FSP11}. This procedure consists of extracting the superconducting
density of states of the tip from spectra on the bare surface and using the
result for fitting the differential conductance spectra of the MnPc spectra
assuming a set of Shiba states (for details see Ref.~\onlinecite{FSP11}, and
the appendix). The result
is representative for the quasi-particle density of states (DOS) of the MnPc
molecule on the superconducting Pb surface (Fig.~\ref{fig:1}(d)). From these plots
we can deduce the energy of the Shiba states and their
intensity. $E_{b,g}$ ($E_{b,s}$) is the energy for the bound state with larger
(smaller) weight $w_{b,g}$ ($w_{b,s}$), where $E_{b,g}=-E_{b,s}$. We observe a
gradual increase in the asymmetry of the weights when $E_{b,g}$ shifts from
negative values to positive ones. This is shown in Fig.~\ref{fig:wbgowbs} 
as ratio $w_{b,g}/w_{b,s}$.

\begin{figure}[!htbp]
%\vspace*{1cm}
\centering
\includegraphics[width=0.95\columnwidth]{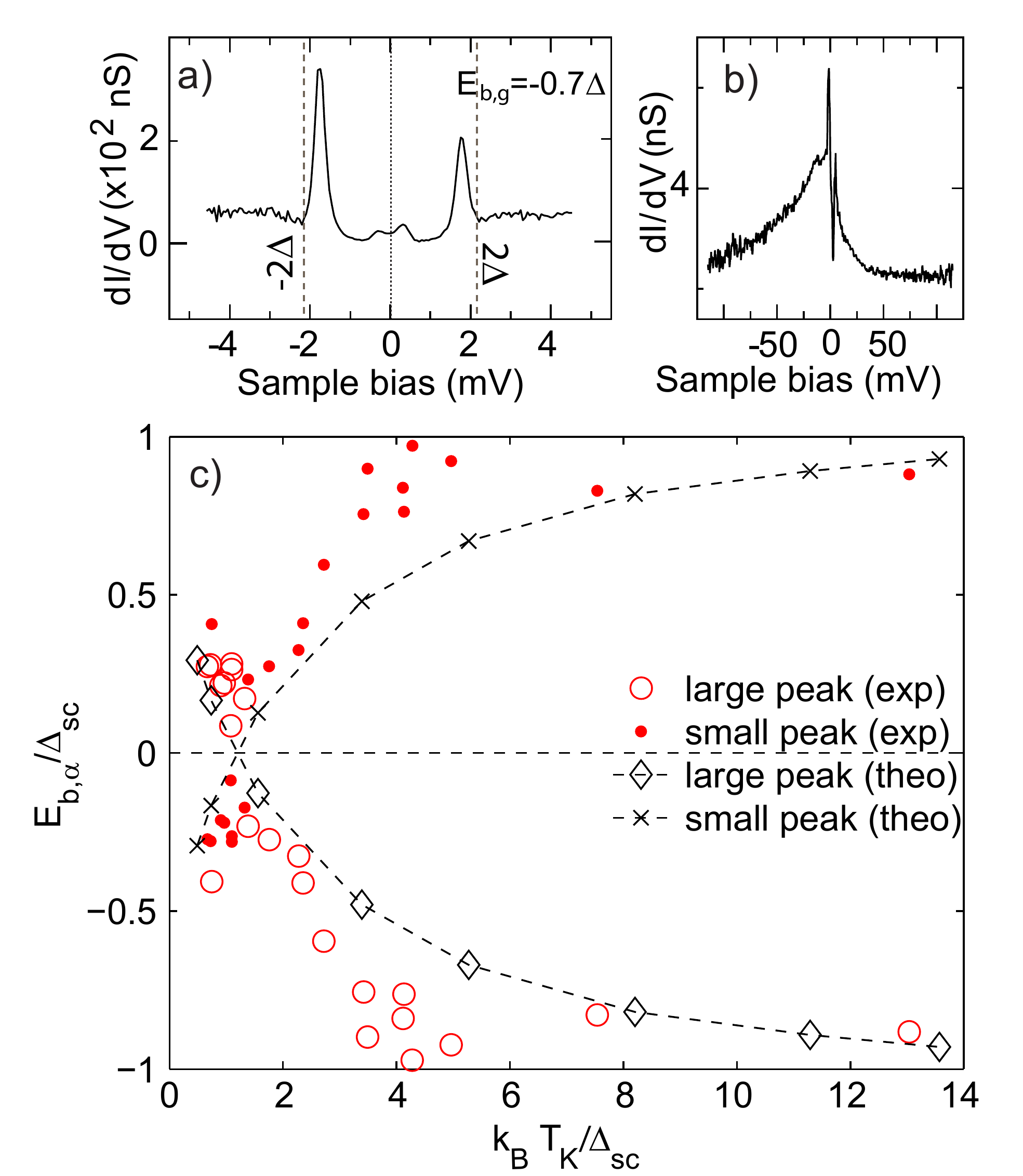}
\vspace*{-0.5cm}
%\vspace*{-0.3cm}
\caption{(Color online) a) Tunneling spectrum showing the Shiba states
  (feedback opened at I\,=\,450\,pA,  $V$\,=\,8.6\,mV). b) Tunneling
  spectrum on the same molecule as in (a) in a larger energy scale
  evidencing a broad background, which can be fitted by two Kondo resonances (feedback opened at I\,=\,470\,pA,
  $V$\,=\,130\,mV). c) Bound state energies $E_{b,\alpha}/\Delta_{\rm sc}$  
  vs Kondo temperature $T_{\rm K}/\Delta_{\rm sc}$. Comparison of
  experimental results with theory for the same choice of model
  parameters as in Fig.~\ref{fig:wbgowbs}.}           
\label{fig:EbvsTK}
\end{figure}

We additionally identify a broad peak at the MnPc center around the Fermi
energy (see Fig.~\ref{fig:EbvsTK}(b)), which can be fitted by two Fano
lineshapes, representing two different Kondo screening processes with $T_{\rm
  K,2}\gg T_{\rm K,1}$.\cite{FSP11} $ T_{\rm K,2}$ with $ \approx 15 -
45$\,meV scales with $ T_{\rm K,1} \approx 1 - 5$\,meV (for details the appendix).  
The occurrence of two Kondo screening channels can be related to the spin state of the MnPc molecule. For the isolated MnPc complex, density functional theory (DFT)
calculations have found that the high spin configuration of Mn
($S=5/2$) is reduced to $S=3/2$ and the unpaired electrons occupy 
the $b_{2g}(d_{xy})$, the $e_g(d_{\pi})$, and the
$a_{1g}(d_{z^2})$ orbital ($x$-axis along an arm of MnPc) and are aligned
due to Hund's rule coupling.\cite{LWH05}
When adsorbed on the Pb surface (in $z$ direction), the
$d_{z^2}$ orbital hybridizes strongly with the Pb states and
is therefore assumed to be quenched.\cite{Fea07} Hence, the observation of two
screening channels suggests a spin state of $S=1$. Since  $k_{\rm  B} T_{\rm 
  K,2}\gg\Delta_{\rm sc}$, Kondo screening dominates for this channel.\cite{BVZ06}   
Therefore, we only correlate $T_{\rm K,1}\equiv T_{\rm K}\sim \Delta_{\rm
  sc}$ with the appearance of the bound states inside the gap.

Fig.~\ref{fig:EbvsTK} shows the dependence of $E_{b,\alpha}$ on $R_{\rm
  K,sc}=k_{\rm B}\, T_{\rm  K}/\Delta_{\rm sc}$. At large $R_{\rm K,sc}$, the
Kondo screening is efficient and the many-body ground state is a singlet.  
Here tunneling can occur via the doublet state which involves breaking
of the Kondo singlet and rearrangement of Cooper pairs.  With
decreasing $R_{\rm K,sc}$ this requires less energy and 
we find $E_{b,\alpha}\to 0$. The level crossing occurs at $R_{\rm K,sc}^c\simeq
1.2$. This point can be regarded as the critical point of the QPT and is a
universal feature. Further decrease of $R_{\rm K,sc}$ leads to the unscreened
doublet ground state.

%\paragraph{Theoretical details -} 
\section{Theoretical results for the one channel model}
For the theoretical description we use an Anderson impurity model (AIM) Hamiltonian of the form    
\begin{equation}
  H=H_{\rm sc}+H_{d}+H_{\rm mix}.
\label{eq:ham}
\end{equation}
The superconducting medium reads,
\begin{equation}
H_{\rm sc}\!= \!\!\!\sum_{\vk,m,{\sigma}} \!\!\!\epsilon_{\vk,m} \elcre{\vk}{m,\sigma}
\elann{m,\vk}{\sigma} -  
\sum_{\vk,m}(\Delta_{\rm sc}\elcre{\vk}{m,\uparrow} \elcre{-\vk}{m,\downarrow}
+\hc)
\end{equation}
where $\elcre{\vk}{m,\sigma}$ creates a band electron with momentum $\vk$,
spin $\sigma$ and band index $m$, where $m=1,\ldots N_c$, and there are
$N_c$ available channels. $\epsilon_{\vk,m}$ is the corresponding electronic
dispersion and $\Delta_{\rm sc}$ the gap parameter chosen real. The band electrons hybridize
with the impurity states via 
\begin{eqnarray}
H_{\rm mix}
=\sum_{\vk,m,{\sigma}}(V_{m}\elcre{\vk}{m\sigma}\elann{d}{m,\sigma} + \hc),
\end{eqnarray}
where $\elcre{d}{m,\sigma}$ creates a d-level impurity electron with
spin $\sigma$ and index $m=1,\ldots N_d$. In the present situation, the number
of conduction channels which hybridize is equal to the d-orbital states,
i.e. $N_d=N_c$.  We will assume different matrix elements 
$V_m$ due to different overlapping integrals. These matrix elements determine
the energy scale for the hybridization of d-states with the substrate through
$\Gamma_m=\pi V_m^2\rho_{m,c}$, where $\rho_{m,c}$ is the DOS of the
conduction band at the Fermi level $\epsilon_{\rm F}$. 
Despite the experimental observation of two Kondo screening channels, we will
now show that the behavior of the Shiba states
can be well described by a single channel model ($N=1$), suggesting a
low energy decoupling of the Kondo channels.  
For the single channel case the ``d-orbital'' term $H_d$ simply reads, 
\begin{equation}
H_{ d}= \sum_{{\sigma}}\epsilon_{d}n_{\sigma} 
 + Un_{\uparrow}n_{\downarrow},
\end{equation}
with the d-level position $\epsilon_{d}$ relative to $\epsilon_{\rm F}=0$
and the on-site Coulomb interaction with 
strength $U$, where $n_{\sigma}=\elcre{d}{\sigma}
\elann{d}{\sigma}$.  For this model we perform NRG calculations \cite{BCP08}
to calculate the lowest energy excitations and their spectral weights, which
characterize the subgap bound states. Examples for the low energy spectra can
be seen in Fig.~\ref{fig:1} (e).

We now explain how to choose the model parameters 
for the different MnPc molecules on the Pb(111) surface. We expect
that the main difference for the MnPc molecules in the different adsorption
sites is the magnitude of the hybridization $\Gamma$, which in turn leads 
to different $T_{\rm K}$. The energy level alignment of the d-states
$\epsilon_{d}$ and Coulomb energy $U$ are, on the 
contrary, expected to change little with the site.\cite{Ji10} We therefore
{\em only} vary $\Gamma$ to explain the data. The superconducting gap
$\Delta_{\rm sc}=1.1$\,meV sets the energy scale. 
The relation of values for $\epsilon_{d}$, $\Gamma$ and $U$ is constrained to
give suitable values for the Kondo temperature $T_{\rm K}\sim \Delta_{\rm
  sc}$. Their actual value can be fixed by matching the strong experimental  
weight asymmetry $w_{b,g}/w_{b,s}$ of the Shiba states for the maximal
$E_{b,g}>0$ on the doublet side in Fig.~\ref{fig:wbgowbs}. This yields
$\epsilon_{d}/\Delta_{\rm sc}\approx -73$, $U/\Delta_{\rm sc}\approx 91$, and
$\Gamma/\Delta_{\rm sc}=16$. This corresponds to an asymmetric AIM,
with $\xi=\epsilon_{d}/U+1/2\approx 0.3$. A variation of $\Gamma/\Delta_{\rm sc}$ from
$16-46$ then reproduces accurately the weight asymmetry in
Fig.~\ref{fig:wbgowbs} on decreasing $E_{b,g}$ and also the variation of
$E_b^{\alpha}$ with $R_{\rm  K,sc}$ in Fig.~\ref{fig:EbvsTK}. 
NRG calculations for an asymmetric one-channel AIM can thus account for the
experimentally observed universal features such as $R_{\rm  K,sc}^c$ and
non-universal ones like $w_{b,g}/w_{b,s}$.\footnote{Other methods
  \cite{FB97a,FB97b,SBS97}  can give asymmetric weights even in more symmetric
  situations, however, they are based  on classical spins, such that the Kondo
  effect is not captured.}

Most important for the understanding of the physical processes is the correct
description of the QPT. For the one channel
non-degenerate AIM and the Kondo model, NRG studies
\cite{SSSS92,SSSS93,YO00,BOH07} have estimated that the phase transition
occurs when $R_{\rm  K,sc}^c\simeq 0.3$. A deviation from this would indicate
that a different number of channels contribute to the Kondo screening of the
same electron spin state.\cite{ZBP11} In particular, this may illustrate the
role of the second Kondo channel observed in the experiment.   
  
At first sight the experimental result for the QPT, $R_{\rm K,sc}^c\simeq 1.2$
(Fig.~\ref{fig:EbvsTK}), seems to suggest a more
complicated situation than a one-channel model. 
However, we find that the origin of this discrepancy is the use of different
definitions of $T_{\rm K}$, which can vary by a prefactor. In the theoretical works
\cite{SSSS92,SSSS93,YO00,BOH07} the definition \cite{Hal78} 
\begin{equation}
    T_{\rm K}=0.29[U\Gamma]^{1/2}\e^{\frac{\pi\epsilon_{\rm d}(\epsilon_{\rm
          d}+U)}{2\Gamma U}}
\label{TK4}
\end{equation}
was used. 
For the experimental values of $T_{\rm K}$, we employ the widely used
definition,\cite{Goldhaber98,Nagaoka02} based on the width of the  
Kondo resonance $\Delta_{\rm K}$ (half width at half maximum) in the
limit $T\to 0$. 
We adopt the same definition in our NRG calculations ($T_{\rm
  K}=\Delta_{\rm K}$) and find that $\Delta_{\rm K}$ and the
definition in Eq.~(\ref{TK4}) can differ by a factor 4 (see also
Ref.~\onlinecite{Cos00}). Taking this into account our result for the QPT in
Fig.~\ref{fig:EbvsTK} is in excellent agreement with the 
theoretical prediction for a one-channel model. We can thus conclude that the
second Kondo screening channel does not shift the transition point.

%\paragraph{Effective model and scaling theory -} 
\section{Derivation of the effective model and scaling theory}
We now discuss the emergence of the low energy effective one-channel model and
its parameters using scaling arguments. First notice that the magnitudes of
$\epsilon_{d}$, $U$ and  
$\Gamma$ do not correspond to usual atomic values $\sim O(eV)$, but
rather to values of the order $100$\,meV. \footnote{Calculations with
  $\epsilon_{d}$, $U$ $\sim O(eV)$ with varying $\Gamma$ do not reproduce the
  behavior of $w_{b,g}/w_{b,s}$ very well.}
This is related to the fact that the AIM under consideration is an effective
model valid for low energies.  
Using insights from the DFT calculations for MnPc,\cite{LWH05} and the
observation of two Kondo channels, we 
start with a model of the form of equation (\ref{eq:ham}) with the two
d-levels coupled to two bands ($N_c$$=$$N_d$$=$$2$) with the hybridization
terms  
$\Gamma_1$$<$$\Gamma_2$. Channel 1 thus describes the physical processes
related to the smaller Kondo temperature.
We assign it to the $d_{xy}$ orbital, since due the spatial orientation the 
overlap with the substrate is much larger for the $d_{\pi}$
orbital.\footnote{The possibility of a different assignment of screening
  channels 
  can not be excluded (see e.g. Ref.~\onlinecite{Fea07}). However, this assignment leads
  to a consistent picture.}   
The impurity term can be written in terms of the level positions
$\epsilon_{d,m}$ relative to $\epsilon_{\rm F}$, intra orbital Coulomb energy $U_m$,
inter orbital Coulomb interaction $U_{12}$ and a Hund's rule 
interaction $J_{\rm H}$,  
\begin{eqnarray}
\label{eq:twochannel}
H_{\rm d}&=& \sum_{{\sigma},m}\epsilon_{d,m}\elcre{d}{m,\sigma}
 \elann{d}{m,\sigma} 
 + \sum_{m} U_m n_{d,m,\uparrow}n_{d,m,\downarrow} \\
&& +\sum_{\sigma,\sigma'}U_{12} n_{d,1,\sigma}n_{d,2,\sigma'}
- J_{\rm H}\vct S_{d,1} \cdot\vct S_{d,2}, \nonumber
\end{eqnarray}
where $S^{\alpha}_{d,m}
=\sum_{\sigma_1,\sigma_2}\elcre{d}{m,\sigma_1}\sigma^{\alpha}_{\sigma_1,\sigma_2}\elann{d}{m,\sigma_2}$
with the Pauli matrix $\sigma^{\alpha}$. The complete set of bare parameters of this
model can not be extracted from existing DFT calculations,\cite{LWH05}
or experimental observations. Therefore, the following are qualitative
arguments on general grounds. 

%\begin{figure}[!htbp]
\begin{figure}[!t]
%\vspace*{1cm}
\centering
\includegraphics[width=0.95\columnwidth]{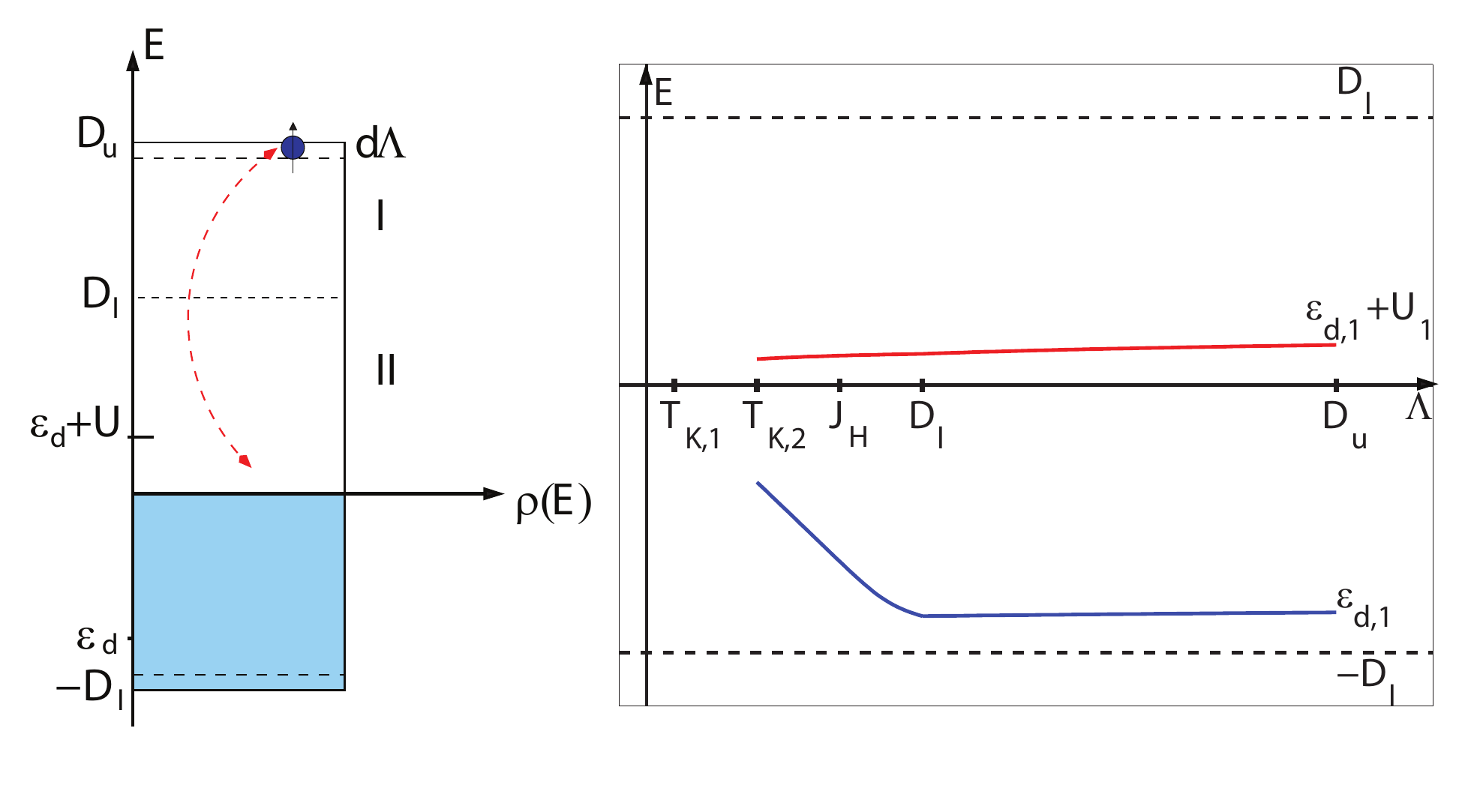}
\vspace*{-0.5cm}
%\vspace*{-0.3cm}
\caption{(Color online) Left: Schematic plot of the bare band and
  impurity energy scales and depiction of two scaling regimes
  (I,II). Right: Schematic plot of the scaling of the effective level
  $\epsilon_{d,1}$ and $\epsilon_{d,1}+U_1$ in the two regimes 
  $\Lambda\in(D_l,D_u)$ and $\Lambda\in(T_{\rm K,2},D_l)$. Scaling
  stops at $\Lambda\sim T_{\rm K,2}$, and the effective one band model is
  used to compute the low energy properties.}         
\label{fig:scaling}
\vspace*{-0.5cm}
\end{figure}
%\noindent

We now derive scaling equations,\cite{And70,Hal78} which connect the original
model and its parameters at a high energy scale $\Lambda$ of the order of the
electronic bandwidth to the low energy effective ones at $\Lambda\sim T_{\rm
  K,2}$, where the second channel decouples due to its complete Kondo
screening. We focus on the quantities in channel 1, where the Shiba states 
occur. 
We only include processes with matrix elements  
$\sim V_1^2$ explicitly, where the level occupations of $\epsilon_{d,1}$
changes. There are also contributions $\sim V_2^2$, whose inclusion
leads to quantitative changes in the equations but does not alter the 
conclusions (for details see the appendix). 
Band structure calculations \cite{ZEP80} for Pb suggest a situation,
where the band $(-D_{l},D_{u})$ is asymmetric with respect to
$\epsilon_{\rm  F}$, $D_{l}<D_{u}$. Then, we have two scaling regimes
(I, II, see Fig. \ref{fig:scaling}): I, where the scaling process has no
counterpart in the occupied states, and II, where the residual band structure is
symmetric around $\epsilon_{\rm  F}$. The scaling equations for the energies
$\epsilon_{d,1}$ and $U$ in regime I, i.e., in $\Lambda\in  (D_{l},D_{u})$
read,\cite{Hal78,hewson}
\begin{eqnarray}
  \frac{d \epsilon_{d,1}}{d\Lambda}&\simeq&\frac{\Gamma_1}{\pi}\frac{1}{\Lambda-\epsilon_{d,1}}, \\
  \frac{d U_1}{d\Lambda}&\simeq&\frac{2\Gamma_1}{\pi}\frac{U_1}{(\Lambda-\epsilon_{d,1})(\Lambda-\bar \epsilon_{d,1})} ,
\end{eqnarray}
where $\bar \epsilon_{d,1}=\epsilon_{d,1}+U_1$.
Different from the usual approaches we also use a scaling equation for
$\Gamma_1$, which is derived from $d T_{\rm   K,1}/d\Lambda=0$,
assuming $T_{\rm K,1}$ as a scaling invariant. Such a scaling
procedure can be continued as long as the levels
$|\epsilon_{d,1}|$, $|\epsilon_{d,1}+U_1|$ lie within $(-\Lambda,\Lambda)$
and do not interfere with the Kondo scale.
We find $\frac{d   \epsilon_{d,1}}{d\Lambda}>0$ and $\frac{d
  U_1}{d\Lambda}>0$, such that $\epsilon_{d,1}<0$ is shifting away from
$\epsilon_F$ when scaling to lower energy and $U$ decreases. The model becomes
more asymmetric in this regime. When $\Lambda\sim D_{l}$, we reach the scaling
regime II, with the corresponding scaling equations  
\begin{eqnarray}
\frac{d \epsilon_{d,1}}{d\Lambda} &\simeq&
\frac{\Gamma_1}{\pi}\frac{2\epsilon_{d,1}(\epsilon_{d,1}+\Lambda)+U_1(3\epsilon_{d,1}-\Lambda)}{(\Lambda^2-\epsilon_{d,1}^2)(\Lambda+\bar
  \epsilon_{d,1})}, \label{eq:scalII1} \\ 
 \frac{d U_1}{d\Lambda} &\simeq&
\frac{4\Gamma_1}{\pi}\frac{\bar
  \epsilon_{d,1}(\Lambda^2-\epsilon_{d,1}^2)-\epsilon_{d,1}(\Lambda^2-\bar
  \epsilon_{d,1}^2)}{(\Lambda^2-\epsilon_{d,1}^2)(\Lambda^2-\bar
  \epsilon_{d,1}^2)}. \label{eq:scalII2}
\end{eqnarray}
For  starting values, $\epsilon_{d,1}(D_{l})<0$, $U_1(D_{l})>0$, 
one has $\frac{d \epsilon_{d,1}}{d\Lambda}<0$. 
In this case $\epsilon_{d,1}$ is shifting towards $\epsilon_F$. Due to the term
in the denominator the effect becomes strong when $\Lambda^2 \sim
\epsilon_{d,1}^2$ [see Fig.~\ref{fig:scaling} (right)]. 
For the same effective starting values we have $\frac{d
  U_1}{d\Lambda}>0$. Hence, $U_1$ and ($\epsilon_{d,1}+U_1)$ decrease
further under the scaling. This scaling can be continued until we
reach $\Lambda\sim J_{\rm H}$. We have  $D_l\sim
3$\,eV, and for usual estimates for $J_{\rm H}$  we expect $T_{\rm
  K,2}<J_{\rm H}<D_{l}$. At this scale the spins lock into the high spin $S=1$
configuration due to the dominating Hund's coupling. As shown in
Ref.~\onlinecite{NC09}, this leads to a reduction of the magnetic coupling $J\to
J/(2S)$ and the Kondo scale. In our approach it can be included in $\Gamma_1$,
such that the scaling is slowed 
down.\cite{NC09} The scaling can then be continued until  
$\Lambda\sim T_{2,\rm   K}$. Then one part of the impurity spin
becomes Kondo screened and decouples.  At this scale we have an effective
single band model with $S=1/2$. Its parameters are
substantially reduced from the bare values and generically 
asymmetric, which clarifies our choice in the earlier NRG calculations. 
Since NRG calculations for multi-channel models are very challenging, our
approach combined with scaling equations could be 
useful in other situations for complex molecules on surfaces.

In conclusion, we provide a unified experimental and theoretical perspective
of the microscopic interplay of Kondo screening and superconducting pairing
for MnPc on lead, as manifested in the subgap bound states. We identify the
change of hybridization as the relevant quantity and show that in spite of the
complex spin state of MnPc and two Kondo screening channels, an effective
description based on the one-channel AIM captures both universal aspects like
$R_{\rm K,sc}^c$ and 
non-universal ones like the asymmetry of the weights very well.  
In the future it would be interesting to envisage situations, where 
different impurity spins can interact, such that different kinds of quantum phase
transitions can occur.  

\bigskip
\noindent{\bf Acknowledgment}\par
\noindent
We wish to thank P. Coleman, M. Haverkort,
A.C. Hewson,  A. Subedi for helpful discussions and G. Schulze for the
deconvolution program of tunneling spectra. JB acknowledges financial
support from the DFG through BA 4371/1-1. We also thank the Focus area
Nanoscale of Freie Universit\"at Berlin and the Deutsche
Forschungsgemeinschaft through Sfb 658 for financial support.  

\begin{appendix}
\section{Deconvolution procedure}

In conventional STM experiments, to first order approximation the $dI/dV$
spectra resemble the density of states of the sample. This relies on a
constant density of states of the STM tip. In the presented experiment, the
tip is coated with the superconducting material of the substrate, thus
exhibiting a Bardeen-Cooper-Schrieffer (BCS) like density of states. To
interpret the measured  $dI/dV$ spectra as the density of states of the sample,
the influence of the tip DOS has to be extracted. We do this by a
deconvolution procedure, which has been described and tested in the
supplementary material of Ref.~\onlinecite{FSP11}. 

The STM current $I$ as a function of voltage $V$ has the following form,
\begin{equation}
 I(V,\vct r)\sim \int d\omega \rho_s(\omega,\vct r) \rho_t(\omega)
 [f(\omega)-f(\omega-eV)], 
\end{equation}
where $\rho_s(\omega,\vct r)$ is the local density of states of the sample at
position $\vct r$, $\rho_t(\omega)$ the one of the tip and $f(\omega)$ is the
Fermi function. 
The deconvolution method consists of a two step process based on
fitting the experimental spectra with a simulated density of states. In the
first step, the experimental spectrum of the clean surface
$\rho_s(\omega,\vct r)=\rho_s(\omega)$ is measured and fitted. The density of
states of sample and 
tip are equal, $\rho_s(\omega)=\rho_t(\omega)$, and follow a BCS density of
states broadened by a Lorentzian function. To reproduce the $dI/dV$ spectra, we
calculate the tunneling current at each bias 
voltage and numerically derive the $dI/dV$ function. This function is then
fitted to the experimental data to extract the temperature, superconducting
gap and broadening.  
Since the same tip and experimental conditions are used to measure the
$dI/dV(V,\vct r=\vct R_{\rm MnPc}) $
spectra on the MnPc molecules, these
parameters remain fixed for modeling the 
tip density of states $\rho_t(\omega)$, which is necessary in the next
step. The Shiba states are simulated by two Lorentz peaks inside a symmetric
gap (modeled by two broadened step functions) around the Fermi level. With
this model density of states, the tunneling current is again calculated at
each bias voltage and the $dI/dV$ signal derived numerically. The fitting
procedure then gives a result for the local $\rho_s(\omega,\vct r=\vct R_{\rm
  MnPc})$ from which  the  position, width, and amplitude of the Shiba states
can be extracted.  

To show the validity and precision of the procedure, we present in Fig. 1c) a
direct comparison of the measured and fitted spectra for a number of examples.

\section{Definition of the Kondo temperature}
For a quantitative comparison between experiment and theory it is important
that the same definition of the Kondo temperature $T_{\rm K}$ is used.
In the literature one can find different definitions for the Kondo temperature
$T_{\rm K}$.\cite{hewson} Theoretically, it is convenient to define it via the
magnetic susceptibility of the impurity in the limit $T\to 0$, $\chi_m=x/(4
T_{\rm K})$ with a suitable prefactor $x$. This quantity is however, not 
experimentally measured in the present context. 
The quantity which is most easily accessible experimentally is the width of
the Kondo resonance $\Delta_{\rm K}$ (half width at half maximum). To make
the comparison of experiment and theory consistent we have used this
definition also in the theory, i.e. $\Delta_{\rm K}=T_{\rm K}$. 
The experimental data is well understood in terms of an overlap of two Fano
functions for the two Kondo channels with two different Kondo temperatures
$T_{\rm K,1}$, $T_{\rm K,2}$.  
For details Ref.~\onlinecite{FSP11} can be consulted. In Table \ref{table:TK1TK2} we
display a number of values found on different molecules.
\begin{table}[!htpb]
\begin{tabular}{| c| c| c |}\hline
 $E_{b,g}/\Delta_{\rm sc}$ & $T_{\rm K,1}$ (K) & $T_{\rm K,2}$ (K) \\ \hline
%-0.86 & 58$\pm$21 & 410$\pm$50 \\ \hline
-0.76 & 52$\pm$5 & 480$\pm$60 \\ \hline
-0.69 & 53$\pm$8 & 330$\pm$40 \\ \hline
-0.46 & 38$\pm$2 & 280$\pm$50 \\ \hline
-0.29 & 29$\pm$5 & 420$\pm$ 30 \\ \hline
 0.24 & 14$\pm$10 & 210$\pm$ 50 \\ \hline
\end{tabular}
 \caption{Exemplary values for $T_{\rm K,1}$ and $T_{\rm K,2}$ on different
   molecules with energy of the larger Shiba state $E_{b,g}/\Delta_{\rm sc}$.}
  \label{table:TK1TK2}
\end{table}

\section{Scaling equations for the two channel model}

Here we present the complete set of scaling equations for the two-channel
model. We will argue that the conclusions in the main text are not changed
qualitatively due to additional terms proportional to $V_2^2$.
The starting point is the effective two orbital model in equation (\ref{eq:ham}) together
with (\ref{eq:twochannel}). In the following we focus on the complete scaling equations in
regime II. The scaling equations in regime I only contain a part of the terms
and show how the asymmetry is increased for an asymmetric band. They can be
derived and discussed in a similar fashion. 
For each orbital there are 4 possible states, such that there are 16 atomic
states which can be described by the occupation numbers 
$n_{1,\sigma}$, $n_{2,\sigma}$ for orbital 1 and 2, respectively.
Due to the Hund's rule term ($J_{\rm H}>0$),
\begin{equation}
  H_H=-J_{\rm H} \vct S_{d,1}\cdot \vct S_{d,2},
\end{equation}
the atomic eigenbasis for the singly occupied situation,
$n_{1,\sigma}=n_{2,\sigma}=1$ is given by singlet and triplet states,
$\ket{S,S_z}$, and hence we use those quantum numbers, where $S$ is the total
spin with $z$-component $S_z$. One has 
\begin{equation}
  H_H \ket{1,S_z}=-J_{\rm H}\ket{1,S_z}, \;\;
  H_H \ket{0,S_z}=3 J_{\rm H}\ket{0,S_z},
\end{equation}
i.e., the triplet state state has the lowest energy, and the energy difference 
is $4J_{\rm H}$ for our definition of $\vct S_{d,i}$, which does not include a
factor $1/2$. For the derivation of the scaling equations it is useful to
define Hubbard operators $X_{ab}=\ket{a}\bra{b}$ for the atomic
states.\cite{hewson} The atomic terms can be given in diagonal energy  
representation with $E_{n_{1,\sigma},n_{2,\sigma}}$ and $\bar E_{S, S_z}$. The
hybridization term constitutes of a lengthy expression in terms of those
Hubbard operators taking into account all possible processes of different
occupation.  
The single particle occupation parameters of the Hamiltonian can be expressed as
$\epsilon_{d,1,\sigma}=E_{n_{1,\sigma}=1,0}-E_{0,0}$. 
By projecting out high energetic particle and hole states in the bath
\cite{hewson} we find the equations, 
\begin{widetext}
\begin{eqnarray}
 \frac{d  \epsilon_{d,1}}{d\Lambda}&=&\frac{\Gamma_1}{\pi}
\Big[\frac{1}{\Lambda-\epsilon_{d,1}}+\frac{1}{\Lambda+\epsilon_{d,1}+U_1}-\frac{2}{\Lambda+\epsilon_{d,1}} 
\Big]  \label{eq:scalinged1compl}\\
&+& \frac{\Gamma_2}{\pi}
\Big[\frac{3/2}{\Lambda+\epsilon_{d,2}+U_{12}-J_{\rm
    H}}+\frac{1/2}{\Lambda+\epsilon_{d,2}+U_{12}+3 J_{\rm H}}-\frac{2}{\Lambda+\epsilon_{d,2}}
\Big], \nonumber
\end{eqnarray}
\begin{eqnarray}
 \frac{d
   U_1}{d\Lambda}&=&\frac{2\Gamma_1}{\pi}\Big[\frac{1}{\Lambda+\epsilon_{d,1}}
 -\frac{1}{\Lambda-\epsilon_{d,1}}+\frac{1}{\Lambda-\epsilon_{d,1}-U_1}
 -\frac{1}{\Lambda+\epsilon_{d,1}+U_1}\Big]  \label{eq:scalingU1compl} \\
&+&\frac{\Gamma_2}{\pi}\Big[\frac{2}{\Lambda+\epsilon_{d,2}+2U_{12}}-\frac{3}{\Lambda+\epsilon_{d,2}+U_{12}-J_{\rm
  H}} -\frac{1}{\Lambda+\epsilon_{d,2}+U_{12}+3J_{\rm
  H}}+\frac{2}{\Lambda+\epsilon_{d,2}} \Big].
\nonumber
\end{eqnarray}
Similar equations are found for $\epsilon_{d,2}$ and $U_2$, where one has to
interchange the indices $\alpha=1,2$ in $\epsilon_{d,\alpha}$,
$\Gamma_{\alpha}$ and $U_{\alpha}$ on the right hand side. An equation for
$J_{\rm H}$ follows from $4J_{\rm   H}=\bar E_{S=0,S_z=0}-\bar E_{S=1,S_z=1}$, 
\begin{eqnarray*}
 \frac{d J_{\rm H}}{d\Lambda}&=&\sum_{m=1}^2\frac{\Gamma_m}{4\pi}
\Big[\frac{1}{\Lambda-(\epsilon_{d,m}+U_{12}+3J_{\rm H})}
 +\frac{1}{\Lambda+\epsilon_{d,m}+U_{12}+U_1-3J_{\rm
     H}} \\
&&-\frac{1}{\Lambda-(\epsilon_{d,m}+U_{12}-J_{\rm H})}
 -\frac{1}{\Lambda+\epsilon_{d,m}+U_{12}+U_1+J_{\rm H}}\Big].
\end{eqnarray*}
We can represents $U_{12}$ as
$U_{12}=\bar E_{S=1,S_z=1}-E_{0,0}-\epsilon_{d,1}-\epsilon_{d,2}+J_{\rm H}$, and
find
\begin{eqnarray*}
 \frac{d U_{12}}{d\Lambda}&=&\sum_{m=1}^2\frac{\Gamma_m}{\pi}
\Big[\frac{3/4}{\Lambda-(\epsilon_{d,m}+U_{12}-J_{\rm H})}
 +\frac{3/4}{\Lambda+\epsilon_{d,m}+U_{12}+U_m+J_{\rm H}}
 +\frac{1/4}{\Lambda-(\epsilon_{d,m}+U_{12}+3J_{\rm H})} \\
&& +\frac{1/4}{\Lambda+\epsilon_{d,m}+U_{12}+U_m-3J_{\rm H}}
-\frac{3/2}{\Lambda+\epsilon_{d,m}+U_{12}-J_{\rm H}}
-\frac{1/2}{\Lambda+\epsilon_{d,m}+U_{12}+J_{\rm H}} \\
&&
-\frac{1}{\Lambda+\epsilon_{d,m}+U_m}-\frac{1}{\Lambda-\epsilon_{d,m}}
+\frac{2}{\Lambda+\epsilon_{d,m}} \Big].
\end{eqnarray*}
\end{widetext}
We can use the Kondo scales in the two channels $T_{{\rm K},m}$ as scaling
invariants. There is an approximate form $T_{{\rm K},m}\simeq T_{{\rm
    K},m}^0\exp(-\alpha J_{\rm H})$ for the situation with
Hund's rule coupling,\cite{NCH10} where $T_{{\rm  K},m}^0$ is the Kondo scale for $J_{\rm
  H}=0$. From this we can also derive scaling equations for $\Gamma_m$.  
If one neglects $J_{\rm H}$ in the denominator as being smaller than the other
scales then one finds $d J_{\rm H}/d \Lambda \simeq 0$, such that
the Hund's coupling varies little.\cite{NC09}

The Eqs.~(\ref{eq:scalII1}) and (\ref{eq:scalII2}) in the main text for regime II correspond to
(\ref{eq:scalinged1compl}), (\ref{eq:scalingU1compl}) without the terms
proportional to $\Gamma_2$. In 
order to have Kondo physics in both channels, we need the atomic ground state
to be one where both orbitals are singly occupied, which implies
$\epsilon_{d,m}+U_{12}<0$ and $\epsilon_{d,m}+U_{12}+U_m>0$. We also expect a
deviation from particle-hole symmetry $\epsilon_{d,m}=-U_{12}-U_m/2$.
In the main text we have argued that from the term proportional to $\Gamma_1$
one finds $\frac{d \epsilon_{d,1}}{d\Lambda}<0$. It is easy to see that for $\Lambda+
\epsilon_{d,2}>0$, $\Lambda +\epsilon_{d,2}+U_{12}-J_{\rm H}>0$ and $\Lambda
+\epsilon_{d,2}+U_{12}-J_{\rm H}>0$ the term proportional to $\Gamma_2$ is
also negative, such that $\frac{d \epsilon_{d,1}}{d\Lambda}<0$ is
reinforced. Similarly, the additional term $\sim \Gamma_2$ in
Eq. (\ref{eq:scalingU1compl}) has the same sign as the first term and therefore 
$\frac{d   U_1}{d\Lambda}>0$ is maintained. Hence, the qualitative picture
remains unchanged, when these additional terms are taken into account. There
will however be quantitative changes in the scaling. 
The quantities in channel 2, $\epsilon_{d,2}$, $U_2$ possess a similar scaling
flow, but are not of major interest here as the corresponding Kondo
temperature $T_{\rm K,2}$ largely exceeds the superconducting gap. 
A more complete discussion of the different regimes of the scaling equations
and the different resulting behavior is beyond the scope of this work, but can
be subject of a separate study. 

\end{appendix}

\end{document}